\documentclass[dvips,12pt,a4paper]{article}
\usepackage{a4wide}
\usepackage{epsfig}
\usepackage{amsmath}
\usepackage{latexsym}
\usepackage{array}
\usepackage{graphicx}

  \newlength{\absize}
\newcommand{\dd}{\mbox{{\rm d}}}

\def\lsim{\mathrel{\rlap{\raise 2.5pt \hbox{$<$}}\lower 2.5pt
\hbox{$\sim$}}}

\allowdisplaybreaks 

\catcode`@=11
\def\citer{\@ifnextchar [{\@tempswatrue\@citexr}{\@tempswafalse\@citexr[]}}

\def\@citexr[#1]#2{\if@filesw\immediate\write\@auxout{\string\citation{#2}}\fi
  \def\@citea{}\@cite{\@for\@citeb:=#2\do
    {\@citea\def\@citea{--\penalty\@m}\@ifundefined
       {b@\@citeb}{{\bf ?}\@warning
       {Citation `\@citeb' on page \thepage \space undefined}}%
\hbox{\csname b@\@citeb\endcsname}}}{#1}}
\catcode`@=12

\begin{document}
  \thispagestyle{empty}
  \pagestyle{empty}
  \renewcommand{\thefootnote}{\fnsymbol{footnote}}
\newpage\normalsize
    \pagestyle{plain}
    \setlength{\baselineskip}{4ex}\par
    \setcounter{footnote}{0}
    \renewcommand{\thefootnote}{\arabic{footnote}}
\newcommand{\preprint}[1]{%
  \begin{flushright}
    \setlength{\baselineskip}{3ex} #1
  \end{flushright}}
\renewcommand{\title}[1]{%
  \begin{center}
    \LARGE #1
  \end{center}\par}
\renewcommand{\author}[1]{%
  \vspace{2ex}
  {\Large
   \begin{center}
     \setlength{\baselineskip}{3ex} #1 \par
   \end{center}}}
\renewcommand{\thanks}[1]{\footnote{#1}}
\renewcommand{\abstract}[1]{%
  \vspace{2ex}
  \normalsize
  \begin{center}
    \centerline{\bf Abstract}\par
    \vspace{2ex}
    \parbox{\absize}{#1\setlength{\baselineskip}{2.5ex}\par}
  \end{center}}

\hyphenation{phenomeno-logy}
\renewcommand{\thefootnote}{\fnsymbol{footnote}}
\vfill
\null
\null
\title{
Identification of indirect new physics effects at $e^+e^-$
colliders: the large extra dimensions case
}
\vfill
\author{
A.A.\ Pankov$^{a,b,}$\footnote{\tt pankov@gstu.gomel.by} 
and N.\ Paver$^{c,}$\footnote{\tt nello.paver@ts.infn.it}
}
\begin{center}
$^{a}$ Pavel Sukhoi Technical University,
     Gomel 246746, Belarus \\
$^{b}$ Abdus Salam ICTP, Strada Costiera 11, 34100 Trieste, Italy \\ 
$^{c}$ University of Trieste  and INFN-Sezione di Trieste, 34100
Trieste, Italy 
\end{center}
\vfill

\abstract{We discuss indirect manifestations of graviton exchange, predicted 
by large extra dimensions, in fermion-pair production at a high-energy
$e^+e^-$ collider. By means of specifically defined asymmetries among
integrated angular distributions, the graviton exchange signal can be 
cleanly distinguished from the effects of either vector-vector contact 
interactions or heavy scalar exchanges. The role of initial electron and 
positron beams polarization is also discussed. 
The method is applied to a quantitative assessment of the sensitivity to
the mass cut-off parameter $M_H$ of the KK graviton tower in the ADD 
scenario, and of the potential identification reach of this mechanism   
obtainable at the currently planned Linear Collider.
}
\vspace*{20mm}
\setcounter{footnote}{0}
\vfill

\newpage
    \setcounter{footnote}{0}
    \renewcommand{\thefootnote}{\arabic{footnote}}

\section{Introduction}\label{sec:I}
Although the Standard Model (SM) of particle physics has been experimentally
verified with impressive confirmations, there are both theoretical belief and
mounting phenomenological evidence that this model cannot be considered as
the ultimate theory of fundamental interactions. Accordingly, there exist  
a variety of proposed new physics (NP) scenarios beyond the SM,  
characterized by different kinds of non-standard dynamics involving new
building blocks and forces mediated by exchanges of new heavy states,
generally with mass scales much greater than $M_{W}$ or $M_{Z}$. Searches for 
such non-standard scenarios are considered as priorities for experiments 
at very high energy accelerators.
\par
Clearly, the direct production of the new heavy objects at the Large Hadron
Collider (LHC) and at the $e^+e^-$ Linear Collider (LC), and the measurement
of their couplings to ordinary matter, would allow the unambiguous
confirmation of a given NP model. One hopes this to be the case of the
supersymmetric (SUSY) extensions to the SM. In many interesting cases, 
however, the threshold for direct production of the new heavy particles 
may be much higher than the machine energy. In this situation, the relevant 
novel interaction can manifest itself only indirectly, {\it via} deviations of 
measured observables from the SM predictions, produced by virtual 
heavy quantum exchange. 
\par 
A convenient theoretical representation of this kind of scenarios is based on 
appropriate {\it effective} local interactions among the familiar 
SM particles, that lead to ``low'' energy expansions of the transition 
amplitudes in inverse powers of the above mentioned very large mass scales. 
Such interaction Hamiltonians are described by specific operators
of increasing dimension and, generally, only the lowest-dimension 
significant operator is retained, assuming the constributions of the 
higher-dimensional ones to be strongly suppressed by the higher inverse 
powers of the large mass scale hence negligible. Furthermore, additional 
criteria and symmetries are imposed in order to fix the phenomenologically 
viable forms of such non-standard effective interactions and limit the 
number of the new coupling constants and mass scales to be experimentally 
determined (or constrained).
\par
Since different non-standard interactions can in principle induce similar
deviations of the cross sections from the SM predictions, it is important to 
define observables designed to {\it identify}, among the possible 
non-standard interactions, the actual source of a deviation, were it   
observed within the experimental accuracy. 
\par
Great attention has been given, in the context of the hierachy problem 
between the Planck and the electroweak mass scales, to scenarios involving 
compactified extra dimensions, and we will here focus on the so-called 
ADD model where the extra spatial dimensions, in which gravity only can 
propagate, are of the ``large'' millimeter size 
\citer{Arkani-Hamed:1998rs,Antoniadis:1998ig}. Specifically, we will 
discuss the possibility of {\it uniquely} distinguishing, in $e^+e^-$ 
annihilation into fermion pairs at the Linear Collider, the effects 
of graviton exchange predicted by this scenario from the, 
in principle competing, new physics scenarios represented by four-fermion 
contact effective interactions. While originating in the context of 
compositeness of quarks and leptons \citer{'tHooft:xb,Ruckl:1983hz}, the 
latter can more generally represent a variety of new interactions, generated 
by exchanges of very heavy new objects such as, e.g., heavy $Z^\prime$, 
leptoquarks, heavy scalars, with masses much larger than the Mandelstam 
variables of the considered process. 
\par
Since, according to the above considerations, the deviations are 
suppressed by powers of the ratio between the process Mandelstam variables 
and the square of the large mass scale characteristic of the considered 
novel interaction, the search of {\it indirect} manifestations of such 
new physics will be favoured by high energies (and luminosities), that 
may increase the signal and therefore allow higher sensitivites. 
\par 
In Ref.~\cite{Osland:2003fn}, a particular combination of integrated cross 
sections, the so-called ``center-edge asymmetry'' $A_{\rm CE}$,  
was shown to provide a simple tool to exploit the spin-2 character of 
graviton exchange in high energy $e^+e^-$ annihilation, and to disentangle 
this effect from vector-vector contact 
interactions.\footnote{An application of this method to lepton-pair 
production at hadron colliders was discussed in 
Ref.~\cite{Dvergsnes:2004tw}.} This method should usefully complement 
the ones based on Monte Carlo best fits \cite{Cullen:2000ef}, or on 
integrated differential cross sections weighted by Legendre polynomials 
\cite{Rizzo:2002pc}.
\par 
Here, we shall propose an analysis based on an extension of the above 
mentioned asymmetry, the ``center-edge-forward-backward asymmetry''  
$A_{\rm CE,FB}$, that should allow the unique {\it identification} of 
graviton exchange {\it vs.} contact four-fermion interactions, and we will 
assess the {\it identification} reach obtainable at the planned Linear 
Collider. On the other hand, it will be found that the new asymmetry can  
be used also to determine the {\it discovery} reach on the contact 
interactions free from contamination of graviton exchange, and therefore to 
extract useful information also on that sector of new physics.  
\par
Specifically, in Sect.~2 the differential cross section and the 
deviations from the SM induced by the above mentioned NP scenarios 
are discussed. In  Sects.~3 and 4 the basic observable asymmetries are 
defined, both for unpolarized and for polarized electron and positron beams.   
The identification reaches on graviton exchange and on four-fermion contact 
interactions are numerically derived for ``standard'' Linear Collider 
parameters in Sect.~5. Finally, a few conclusive remarks are given in 
Sect.~6.
\section{Differential cross sections}
\label{sec:II}
We consider the process (with $f\ne e,t$)
\begin{equation}
e^++e^-\to f+\bar{f} \label{proc} \end{equation}
with unpolarized $e^+e^-$ beams. Neglecting all fermion masses with respect
to the c.m. energy $\sqrt s$, the differential cross section can be written
as \cite{Schrempp:1987zy}:
\begin{equation}
\frac{\dd\sigma}{\dd z}=\frac{1}{4}\left(
\frac{\dd\sigma_{\rm LL}}{\dd z}+
\frac{\dd\sigma_{\rm RR}}{\dd z}+
\frac{\dd\sigma_{\rm LR}}{\dd z}+
\frac{\dd\sigma_{\rm RL}}{\dd z}\right).
\label{crossdif}
\end{equation}
Here, $z\equiv\cos\theta$ with $\theta$ the angle between the
incoming electron and the outgoing fermion in the c.m. frame, and
$\dd\sigma_{\alpha\beta}/{\dd\cos\theta}$ ($\alpha,\beta={\rm L,R}$) are the
helicity cross sections
\begin{equation}
\frac{\dd\sigma_{\alpha\beta}}{\dd z}=N_C\frac{3}{8}\sigma_{\rm pt}
\vert {\cal M}_{\alpha\beta}\vert^2\, (1\pm z)^2.
\label{helcross}
\end{equation}
Conventions are such that the subscripts $\alpha$ and $\beta$ in the reduced 
helicity amplitudes ${\cal M}_{\alpha\beta}$ indicate the helicities of the
initial and final fermions, respectively. In Eq.~(\ref{helcross}), the
`$+$' sign applies to the combinations ${\rm LL}$ and ${\rm RR}$, while 
the `$-$' sign applies to the ${\rm LR}$ and ${\rm RL}$ cases. Also,
$\sigma_{\rm pt}=4\pi\alpha^2_{\rm e.m.}/3s$, and
the color factor $N_C\simeq 3(1+\alpha_s/\pi)$ is needed only in the
case of quark-antiquark final states.
\par
In the SM the helicity amplitudes, representing the familiar $s$-channel 
photon and $Z$ exchanges, are given by
\begin{equation}
{\cal M}_{\alpha\beta}^{\rm SM}=Q_e Q_f+g_\alpha^e\,g_\beta^f\,\chi_Z,
\label{smamplit}
\end{equation}
where $\chi_Z=s/(s-M^2_Z+iM_Z\Gamma_Z)\simeq s/(s-M^2_Z)$ for
$\sqrt s\gg M_Z$;
$g_{\rm L}^f=(I_{3L}^f-Q_f s_W^2)/s_W c_W$ and $g_{\rm R}^f=-Q_f
s_W/c_W$ are the SM left- and right-handed fermion couplings to the $Z$, 
with $s_W^2=1-c_W^2\equiv \sin^2\theta_W$; $Q_e$ and $Q_f$ are the
initial and final fermion electric charges. The SM
differential cross section can be decomposed into $z$-even and
$z$-odd parts:
\begin{equation}
\frac{\dd\sigma^{\rm SM}}{\dd z}=\frac{3}{8}\sigma^{\rm SM}
\left(1+z^2\right)+\sigma_{\rm FB}^{\rm SM}z,
\label{dsigmasm}
\end{equation}
where
\begin{equation}
\sigma=\int_{-1}^{1}\frac{\dd\sigma}{\dd z}\,{\dd z}\qquad
{\rm and} \qquad
\sigma_{\rm FB}\equiv \sigma A_{\rm FB}=\left(\int_{0}^{1}-\int_{-1}^{0}\right)
\frac{\dd\sigma}{\dd z}\,{\dd z}
\label{sigmatot}
\end{equation}
denote the total and the forward-backward cross sections, respectively. In
particular, in terms of the amplitudes ${\cal M}^{\rm SM}_{\alpha\beta}$:
\begin{equation}
A_{\rm FB}^{\rm SM}=\frac{3}{4}\hskip 2pt
\frac{\left({\cal M}_{\rm LL}^{\rm SM}\right)^2+
\left({\cal M}_{\rm RR}^{\rm SM}\right)^2 -
\left({\cal M}_{\rm LR}^{\rm SM}\right)^2 -
\left({\cal M}_{\rm RL}^{\rm SM}\right)^2}
{\left({\cal M}_{\rm LL}^{\rm SM}\right)^2+
\left({\cal M}_{\rm RR}^{\rm SM}\right)^2 +
\left({\cal M}_{\rm LR}^{\rm SM}\right)^2 +
\left({\cal M}_{\rm RL}^{\rm SM}\right)^2}.
\label{afbsm}
\end{equation}
\par
Rather generally, in the presence of non-standard interactions coming from 
the new, TeV-scale physics, the reduced helicity amplitudes can be expanded
into the SM part plus a deviation depending on the considered
NP model:
\begin{equation}
{\cal M}_{\alpha\beta}={\cal M}_{\alpha\beta}^{\rm SM}
+\Delta_{\alpha\beta},
\label{amplit}
\end{equation}
where the quantities $\Delta_{\alpha\beta}\equiv\Delta_{\alpha\beta}(\rm NP)$
represent the contribution of the new interaction. The typical examples 
relevant to our discussion are the following ones. 
\medskip
\par
{\bf a)} {\it The large extra dimension scenario}, with exchange of
KK towers of gravitons \citer{Arkani-Hamed:1998rs,Antoniadis:1998ig}. In 
this new physics scenario, gravity propagates
in two or more extra spatial dimensions, compactified to a size $R_c$ of the
millimeter order.\footnote{Actually, two extra dimensions for the 
ADD scenario is close to bounds set by gravitational and 
cosmological experiments.} 
In four dimensions, this translates to a tower of
Kaluza-Klein (KK) graviton states with evenly spaced (and almost continuous)
mass spectrum $m_{\vec n}=\sqrt{{\vec n}^2/R_c^2}$, where $\vec n$ labels 
the KK states. The Feynman rules for KK vertices were given in 
Refs.~\cite{Giudice:1998ck,Han:1998sg}. The exchange of such an object is 
described by a dimension-8 effective Lagrangian, and we here choose the form 
\cite{Hewett:1998sn}:
\begin{equation}
{\cal L}^{\rm ADD}=\frac{4\lambda}{M_H^4}T^{\mu\nu}T_{\mu\nu}.
\label{dim-8}
\end{equation}
In Eq.~(\ref{dim-8}), $T_{\mu\nu}$ is the stress-energy tensor of the SM 
particles, $M_H$ is a cut-off on the summation over the KK spectrum, 
expected to be in the TeV range, and $\lambda=\pm1$. The corresponding 
deviations of the helicity amplitudes for the $e^+e^-$ annihilation process 
(\ref{proc}) under consideration, defined in Eq.~(\ref{amplit}), can be 
written as 
\begin{equation}
\Delta_{\rm LL}({\rm ADD})=\Delta_{\rm RR}({\rm ADD})=f_G(1-2z),\quad
\Delta_{\rm LR}({\rm ADD})=\Delta_{\rm RL}({\rm ADD})=-f_G(1+2z),
\label{ADD}
\end{equation}
where $f_G=\lambda\,s^2/(4\pi\alpha_{\rm e.m.}M_H^4)$ parametrizes the strength
associated with massive, spin-2, graviton exchange.
\par 
Concerning the current experimental limits on $M_H$, from the non-observation 
of deviations from graviton exchange at LEP2 the strongest limits are 
$M_H > 1.20$ TeV for $\lambda = +1$ and $M_H > 1.09$ TeV for 
$\lambda = -1$ \cite{Ask:2004dv}. In hadron-hadron collisions, virtual 
graviton exchange modifies the di-lepton and di-photon production cross 
sections. The combined limit obtained by the CDF and D0 Collaborations at the 
Tevatron Run II $p\bar{p}$ collider is: $M_H > 1.28\hskip 2pt{\rm TeV}$ 
\cite{Unel:2004fn,Gallas:2004im}. Experiments at the LHC are expected to be 
able to explore extra dimensions up to multi-TeV scales 
\cite{Cheung:2004ab,Cheung:1999wt}.
\medskip
\par
{\bf b)} {\it Contact interactions}, such as the vector-vector ones 
envisaged in composite models \citer{'tHooft:xb,Ruckl:1983hz}. These are 
described by the leading dimension-6 operators:
\begin{equation}
{\cal L}_{\rm CI}
=4\pi \sum_{\alpha,\beta}\hskip 2pt
\frac{\eta_{\alpha\beta}}{\Lambda^2_{\alpha\beta}}
\left(\bar e_\alpha\gamma_\mu e_\alpha\right)
\left(\bar f_\beta\gamma^\mu f_\beta\right),
\label{lagra}
\end{equation}
where $\Lambda$s are compositeness mass scales and
$\eta_{\alpha\beta}=\pm1, 0$. Accordingly:
\begin{equation} \Delta_{\alpha\beta}({\rm CI})=\pm\frac{s}{\alpha_{\rm e.m.}}
\frac{1}{\Lambda^2_{\alpha\beta}}.
\label{CI}
\end{equation}
\par
Current limits on $\Lambda$s for the specific helicity combinations 
are model-dependent and significantly vary according the the process 
studied and the kind of analysis performed there. In general, results 
from global analyses are given, and the lower limits are of the order of 
$10\hskip 2pt{\rm TeV}$. A detailed presentation of the situation can be 
found in the listings of Ref.~\cite{Eidelman}.
\par 
As previously mentioned, other new physics scenarios can in principle 
mimic the virtual effects of massive graviton exchange as well as those of 
contact interactions, {\it via} the produced deviations of cross sections 
from the SM predictions. As a representative example, we choose here to 
discuss the case of a heavy scalar exchange in the $t$-channel of process 
(\ref{proc}) with $f=\mu$ and $\tau$, such as the   
sneutrino $\tilde\nu$ relevant to R-parity breaking SUSY interactions 
\cite{Kalinowski:1997bc,Rizzo:1998vf}. In this case, the additional 
contributions to SM helicity amplitudes in Eq.~(\ref{amplit}) can be written 
in the form:   
\begin{equation}
\Delta_{\rm LL}(\tilde\nu)=\Delta_{\rm RR}(\tilde\nu)=0 \qquad 
\Delta_{\rm RL}(\tilde\nu)=\Delta_{\rm LR}(\tilde\nu)=
\frac{1}{2}C_{\tilde\nu}\frac{s}{t-m^2}.
\label{nutilde}
\end{equation}
Here, $m$ is the sneutrino mass, 
$C_{\tilde\nu}=\lambda^2/4\pi\alpha_{\rm e.m.}$ with $\lambda$ a Yukawa 
coupling, and $t=-s(1-z)/2$.\footnote{In principle, a situation where 
there are sneutrino exchanges in both the $s$- and $t$-channels may 
occur \cite{Kalinowski:1997bc,Rizzo:1998vf}.} According to 
Eq.~(\ref{nutilde}), the LL and RR amplitudes are completely free 
from $\tilde\nu$-exchange, whereas the RL and LR ones {\it are} affected. 
Indeed, in the heavy sneutrino limit 
${s,-t\ll m^2}$ assumed here, the deviations in leading order behave 
like the contact four-fermion interaction ones in Eq.~(\ref{CI}) and, in 
particular, are $z$-independent. Only at the next order the propagator effects 
induced by $t/m^2$ can introduce a (expectedly suppressed) angular 
dependence of the deviations analogous to Eq.~(\ref{ADD}). 
\par 
Restrictions from low-energy experiments can be summarized by the inequality  
$\lambda\leq 0.1\times\left({m}/200 \hskip 2pt{\rm GeV}\right)$ 
\cite{Kalinowski:1997bc}. Considering the equivalence 
${m}/\lambda\sim\Lambda/\sqrt{8\pi}$ obtained by comparing 
Eqs.~(\ref{nutilde}) and (\ref{CI}) in the contact-interaction limit, 
current limits on $\Lambda$s in the lepton sector of the order of 10 TeV,  
as obtained from LEP2, can be translated to 
${m}> 2\hskip 2pt \lambda\hskip 2pt{\rm TeV}$.  

\section{Center--edge asymmetries}
\subsection{The center--edge asymmetry {\boldmath $A_{\rm CE}$}}
We define the generalized center--edge asymmetry $A_{\rm CE}$ as
\cite{Osland:2003fn,Pankov:1997da}:
\begin{equation}
A_{\rm CE}=\frac{\sigma_{\rm CE}}{\sigma},
\label{ace}
\end{equation}
where $\sigma_{\rm CE}$ is the difference between the ``central'' and
``edge'' parts of the cross section, with $0<z^*<1$:
\begin{equation}
\sigma_{\rm CE}(z^*)=\left[\int_{-z^*}^{z^*}-
\left(\int_{-1}^{-z^*}+\int_{z^*}^{1}\right)\right]
\frac{\dd\sigma}{\dd z}\,{\dd z}.
\label{sce}
\end{equation}
The integration range relevant to Eq.~(\ref{sce}) is depicted in Fig.~1.
\begin{figure} [h]
\begin{center}
\vspace{-8cm}
\epsfig{file=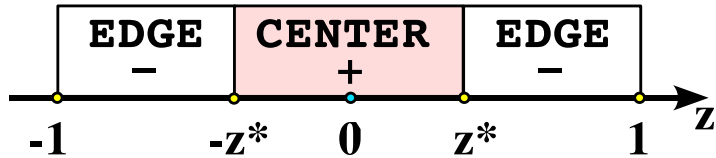,width=15.cm}
\vspace{-8.5cm}
\caption{The
kinematical range of $z\equiv\cos\theta$ and the three bins,
one center and two edge ones, used in definition of  center-edge asymmetry
$A_{\rm CE}$.
}
\end{center}
\end{figure}
Clearly, from the definition Eq.~(\ref{sce}), $\sigma_{\rm CE}=\mp\sigma$
at $z^*=0$ and $1$, respectively. Furthermore, $z$-odd terms in the
differential cross section cannot contribute to $A_{\rm CE}$.
\par
Using Eq.~(\ref{dsigmasm}), one immediately obtains in the SM:
\begin{equation}
A_{\rm CE}^{\rm SM}(z^*)=\frac{1}{2}z^*\hskip 2pt (z^{*2}+3)-1,
\label{acesm}
\end{equation}
independent of the c.m. energy $\sqrt s$, of the flavour of final fermions 
and of the longitudinal beams polarization (which will be considered later).
\par
From the decomposition (\ref{amplit}), one can write $A_{\rm CE}(z^*)$ as 
follows:
\begin{equation}
A_{\rm CE}=\frac{\sigma_{\rm CE}^{\rm SM}+ \sigma_{\rm CE}^{\rm INT}
+\sigma_{\rm CE}^{\rm NP}}
{\sigma^{\rm SM}+\sigma^{\rm INT}+\sigma^{\rm NP}},
\label{acenp}
\end{equation}
where ``SM'', ``INT'' and ``NP'' refer to ``Standard Model'', ``Interference''
and (pure) ``New Physics'' contributions.
\par 
One can quite easily verify, from Eq.~(\ref{CI}), that in the case
where current-current contact interactions are present in addition to the
SM ones, the $z$-dependence of the resulting differential cross section
has exactly the same structure as that of Eq.~(\ref{dsigmasm}), up 
to the superscripts replacement ${\rm SM}\to{\rm CI}$. Consequently,
$A_{\rm CE}$ in this case is identical to the SM one, {\it i.e.},
to Eq.~(\ref{acesm}):
\begin{equation}
A_{\rm CE}^{\rm CI}(z^*)=\frac{1}{2}\,z^*\,({z^*}^2+3)-1.
\label{ACECI}
\end{equation}
Introducing, in general, the deviation of $A_{\rm CE}$ from the SM
prediction:
\begin{equation}
\Delta A_{\rm CE}=A_{\rm CE}-A_{\rm CE}^{\rm SM},
\label{deviat-ace}
\end{equation}
in the case of ``conventional'' contact interactions:
\begin{equation}
\Delta A_{\rm CE}^{\rm CI}=0,
\label{deviat-ace-CI}
\end{equation}
 for any value of $z^*$. Correspondigly, such interactions are 
``filtered out'' by the asymmetry (\ref{ace}), in the
sense that they produce {\it no} deviation from the SM prediction. One can 
easily see that this is the reflection of the postulated vector-vector 
character of Eq.~(\ref{lagra}) and the consequent $z$-independence of the 
right-hand side of Eq.~(\ref{CI}).
\par
Furthermore, Eqs.~(\ref{acesm}) and (\ref{ACECI}) show that
$A_{\rm CE}^{\rm SM}=A_{\rm CE}^{\rm CI}$ vanish
\cite{Osland:2003fn,Datta:2002tk}, at
\begin{equation}
z^*_0=(\sqrt{2}+1)^{1/3}-(\sqrt{2}-1)^{1/3}=0.596,
\label{z*0}
\end{equation}
corresponding to $\theta=53.4^\circ$.
\begin{figure}[t]
\begin{center}
\epsfig{file=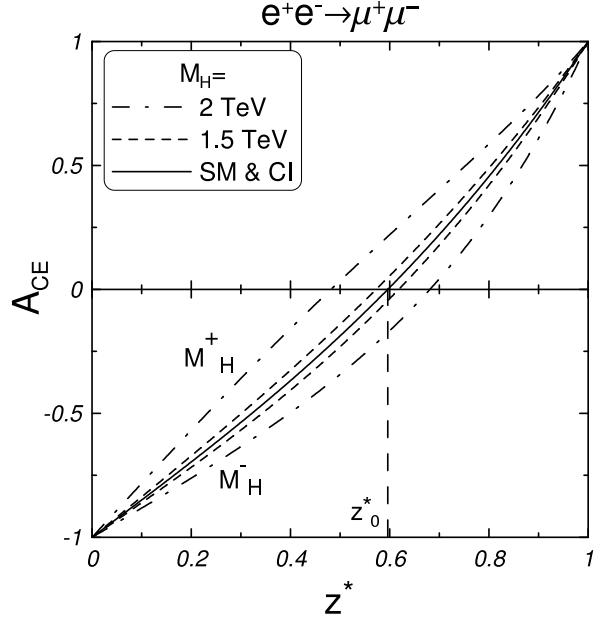,width=9.cm}
\vspace{-3.cm}
\caption{Center-edge asymmetry $A_{\rm CE}$ for $e^+e^-\to\mu^+\mu^-$ 
{\it vs.} $z^*$ in the SM and with CI (solid line) and in the {\rm ADD} 
scenario with $M_H^\pm=1$ TeV (dot-dashed lines) and $M_H^\pm=1.5$ TeV (dashed
lines). The $\pm$ superscripts correspond to constructive and destructive
interference of the SM amplitudes with the graviton exchange.
}
\end{center}
\end{figure}
\par
Spin-2 KK graviton exchange provides characteristic $z$-dependent
deviations of helicity amplitudes, see Eq.~(\ref{ADD}), and {\it non-zero}
values of $\Delta A_{\rm CE}$. In addition, for this kind of new interaction,
$\sigma^{\rm INT}=0$ in the denominator of Eq.~(\ref{acenp}) and the
pure NP contributions proportional to $f_G^2$ should be strongly suppressed
by the high power $(\sqrt s/M_H)^8$, where $(\sqrt s/M_H)^4$ is assumed much
smaller than unity. Therefore, a linear (in $f_G$) approximation to
Eq.~(\ref{acenp}) should numerically be a good approximation and, accordingly, 
one readily derives the expression:
\begin{equation}
\Delta A_{\rm CE}^{\rm ADD}(z^*)\cong \frac{3}{4}f_G\,\frac{
 {\cal M}_{\rm LL}^{\rm SM}+{\cal M}_{\rm RR}^{\rm SM}
-{\cal M}_{\rm LR}^{\rm SM}-{\cal M}_{\rm RL}^{\rm SM}}
{\left[({\cal M}_{\rm LL}^{\rm SM})^2+({\cal M}_{\rm RR}^{\rm SM})^2
+({\cal M}_{\rm LR}^{\rm SM})^2+({\cal M}_{\rm RL}^{\rm SM})^2\right]}\,
4\, z^*\left(1-z^{*2}\right).
\label{deladd}
\end{equation}
\par
In Fig.~2, the $z^*$ dependence of $A_{\rm CE}$ is shown either for
the SM or the CI models and for the ADD scenario (for particular values of
$M_H$). One can conclude that the non-vanishing,
$\Delta A_{\rm CE}\neq 0$, at arbitrary values of $z^*$
(except $z^*=0,1$), or even $A_{\rm CE}\neq 0$
itself for $z^*$ in a range around $z^*_0$, unambiguously signal the
presence of new physics {\it different} from contact four-fermion interactions.
\par
These considerations have been used in Ref.~\cite{Osland:2003fn} to assess the
identification reach on the ADD graviton exchange process with the result
that, depending on the Linear Collider parameters (such as c.m. energy,
luminosity, beams polarizations) and on the final states considered,
the potentially reachable sensitivities to the values of the scale
parameter $M_H$ are in the range of 4--5 TeV at the planned LC with
$\sqrt s = 500$ GeV \cite{Aguilar-Saavedra:2001rg}, up to 10--15 TeV
at a CLIC with $\sqrt s=3-5$ TeV \cite{Assmann:2000hg}.

\subsection{The center--edge--forward--backward asymmetry
{\boldmath $A_{\rm CE,FB}$}}
Still with unpolarized beams, we define this asymmetry as follows:
\begin{equation}
A_{\rm CE,FB}=\frac{\sigma_{\rm CE,FB}}{\sigma},
\label{acefb}
\end{equation}
where, with $0<z^*<1$:
\begin{equation}
\sigma_{\rm CE,FB}=\left[\left(\int_0^{z^*}-\int_{-z^*}^0\right) \,
- \, \left(\int_{z^*}^1-\int_{-1}^{-z^*}\right)\right]\hskip 2pt
\frac{\dd\sigma}{\dd z}\hskip 2pt {\dd z}
\equiv\left(\sigma_{\rm C,FB}-\sigma_{\rm E,FB}\right).
\label{scefb}
\end{equation}
For illustrative purposes, the integration range relevant to Eq.~(\ref{scefb})
is shown in Fig.~3.
\begin{figure} [htb]
\begin{center}
\vspace{-8.cm}
\epsfig{file=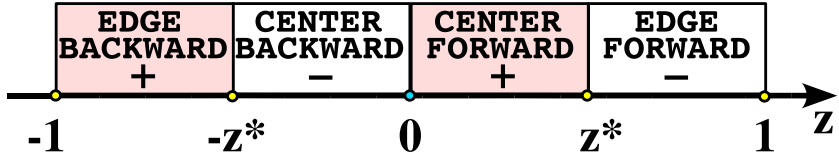,width=15.cm}
\vspace{-8.5cm}
\caption{
The kinematical range of $z$ and the four bins used in the definition of the
center-edge-forward-backward asymmetry $A_{\rm CE,FB}$.
}
\end{center}
\end{figure}
Clearly, $z$-even terms in the differential cross section do not contribute to
Eq.~(\ref{acefb}). Also, by definition, $A_{\rm CE,FB}(z^*=1)=A_{\rm FB}$
and $A_{\rm CE,FB}(z^*=0)= - A_{\rm FB}$.
\par
In the SM, using  Eq.~(\ref{dsigmasm}) one immediately derives
\begin{equation}
A_{\rm CE,FB}^{\rm SM}(z^*)=A_{\rm FB}^{\rm SM}\,(2{z^*}^2-1),
\label{Acefbsm}
\end{equation}
where the expression of $A_{\rm FB}^{\rm SM}$ in terms of the helicity 
amplitudes is given in Eq.~(\ref{afbsm}).
\par
One can immediately see from Eq.~(\ref{CI}) that in the case of the
``conventional'' current-current contact interactions, due to the fact
that the $z$-dependence of the differential cross section remains the
same as in Eq.~(\ref{dsigmasm}) when these interactions add to the SM,
the $z^*$-dependence of $A_{\rm CE,FB}$ will be identical to
Eq.~(\ref{Acefbsm}). Namely:
\begin{equation}
A_{\rm CE,FB}^{\rm CI}(z^*)= A_{\rm FB}^{\rm CI}\,(2{z^*}^2-1).
\label{acefbci}
\end{equation}
Furthermore, Eqs.~(\ref{Acefbsm}) and (\ref{acefbci}) show that both
$A_{\rm CE,FB}^{\rm SM}$ and $A_{\rm CE,FB}^{\rm CI}$ vanish at
\begin{equation}
z^*\equiv z^*_{\rm CI}=1/\sqrt{2}\approx 0.707,
\label{z*CI}
\end{equation}
corresponding to $\theta=45^\circ$.
\par
Stated differently, introducing the deviation of $A_{\rm CE,FB}$ from the SM
prediction:
\begin{equation}
\Delta A_{\rm CE,FB}=A_{\rm CE,FB}-A_{\rm CE,FB}^{\rm SM},
\label{deviat-acefb}
\end{equation}
in the case of ``conventional'' four-fermion contact interactions we have
\begin{equation}
\Delta A_{\rm CE,FB}^{\rm CI}(z^*)=\Delta A_{\rm FB}^{\rm CI}
\left(2{z^*}^2-1\right),
\label{deviat-acefb-ci}
\end{equation}
and the expression of $A_{\rm FB}^{\rm CI}$ in terms of helicity
amplitudes is formally obtained from Eq.~(\ref{afbsm}) by replacing
superscripts ${\rm SM}\to{\rm CI}$. Furthermore:
\begin{equation}
A_{\rm CE,FB}^{\rm SM}(z^*_{\rm CI})=A_{\rm CE,FB}^{\rm CI}(z^*_{\rm CI})=
\Delta A_{\rm CE,FB}^{\rm CI}(z^*_{\rm CI})=0.
\label{zstarci}
\end{equation}
Correspondingly, four-fermion contact interactions are ``filtered out'' also 
by the observable (\ref{acefb}), when measured at $z^*=z^*_{\rm CI}$. One 
can conclude that $A_{\rm CE,FB}\neq 0$ at $z^*_{\rm CI}$ 
unambiguously signals the presence of new physics {\it different} from contact 
interactions. In Fig.~4, we represent the $z^*$-behaviour of 
$A_{\rm CE,FB}^{\rm CI}$ taking in Eq.~(\ref{CI}), for illustrative purposes, 
only the ${\rm LL}$ among the $\eta_{\alpha\beta}$ as a nonvanishing 
CI parameter.

\begin{figure}[htb]
\begin{center}
\epsfig{file=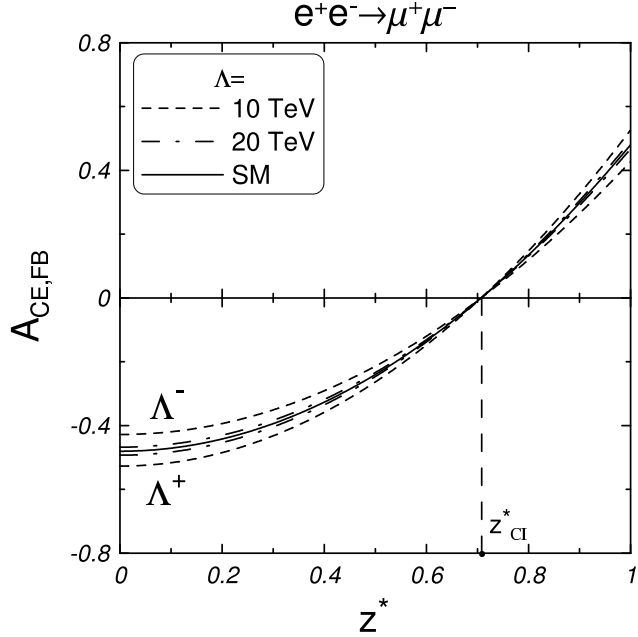,width=10cm}
\vspace{-4.cm}
\caption{
Center-edge-forward-backward asymmetry $A_{\rm CE,FB}^{\rm
CI}$ as a function
of $z^*$ in the SM (solid line) and in the CI model with $\Lambda^\pm=10$
TeV (dashed lines) and $\Lambda^\pm=20$ TeV (dash-dotted lines). The
$\pm$ superscripts correspond to the constructive and destructive 
interference of the SM amplitudes with the CI ones.
}
\end{center}
\end{figure}
\par
Turning to the graviton exchange interaction, the analogue of
Eq.~(\ref{acenp}) is
\begin{equation}
A_{\rm CE,FB}=\frac{\sigma_{\rm CE,FB}^{\rm SM}+ \sigma_{\rm CE,FB}^{\rm INT}
+\sigma_{\rm CE,FB}^{\rm NP}}
{\sigma^{\rm SM}+\sigma^{\rm INT}+\sigma^{\rm NP}},
\label{acefbnp}
\end{equation}
with the same meaning of the superscripts ``SM'', ``INT'' and ``NP'', and
in this case, using Eq.~(\ref{ADD}), one can easily see that
$\sigma^{\rm INT}=0$ {\it and} $\sigma_{\rm CE,FB}^{\rm NP}=0$.
\par
Thus, for the new physics represented by the KK graviton exchange model,
in the approximation of retaining only terms linear in $f_G$ and taking into
account Eq.~(\ref{Acefbsm}), one finds for
the deviation (\ref{deviat-acefb}) the following expression:
\begin{equation}
\Delta A_{\rm CE,FB}^{\rm ADD}(z^*)\cong \Delta A_{\rm FB}^{\rm ADD}
\left(2{z^*}^4-1\right),
\label{deviat-acefb-add}
\end{equation}
where
\begin{equation}
\Delta A_{\rm FB}^{\rm ADD}= -\frac{3}{4}\hskip 2pt f_G\hskip 2pt
\frac{{\cal M}_{\rm LL}^{\rm SM}+{\cal M}_{\rm RR}^{\rm SM}
+{\cal M}_{\rm LR}^{\rm SM}+{\cal M}_{\rm RL}^{\rm SM}}
{\left({\cal M}_{\rm LL}^{\rm SM}\right)^2+
\left({\cal M}_{\rm RR}^{\rm SM}\right)^2+
\left({\cal M}_{\rm LR}^{\rm SM}\right)^2+
\left({\cal M}_{\rm RL}^{\rm SM}\right)^2}.
\label{deviat-afb-add}
\end{equation}
Eq.~(\ref{deviat-acefb-add}) shows that the deviation of $A_{\rm CE,FB}$
from the SM prediction vanishes at 
\begin{equation}
z^*_{\rm G}= 1/2^{1/4}\simeq 0.841,
\label{z*G}
\end{equation}
corresponding to $\theta=33^\circ$, i.e.,
$\Delta A_{\rm CE,FB}^{\rm ADD}(z^*_{\rm G})=0$.
In Fig.~5 we show the $z^*$ behaviour of $A_{\rm CE,FB}^{\rm ADD}$
for selected values of the cut-off parameter $M_H$.

\begin{figure}[htb]
\begin{center}
\epsfig{file=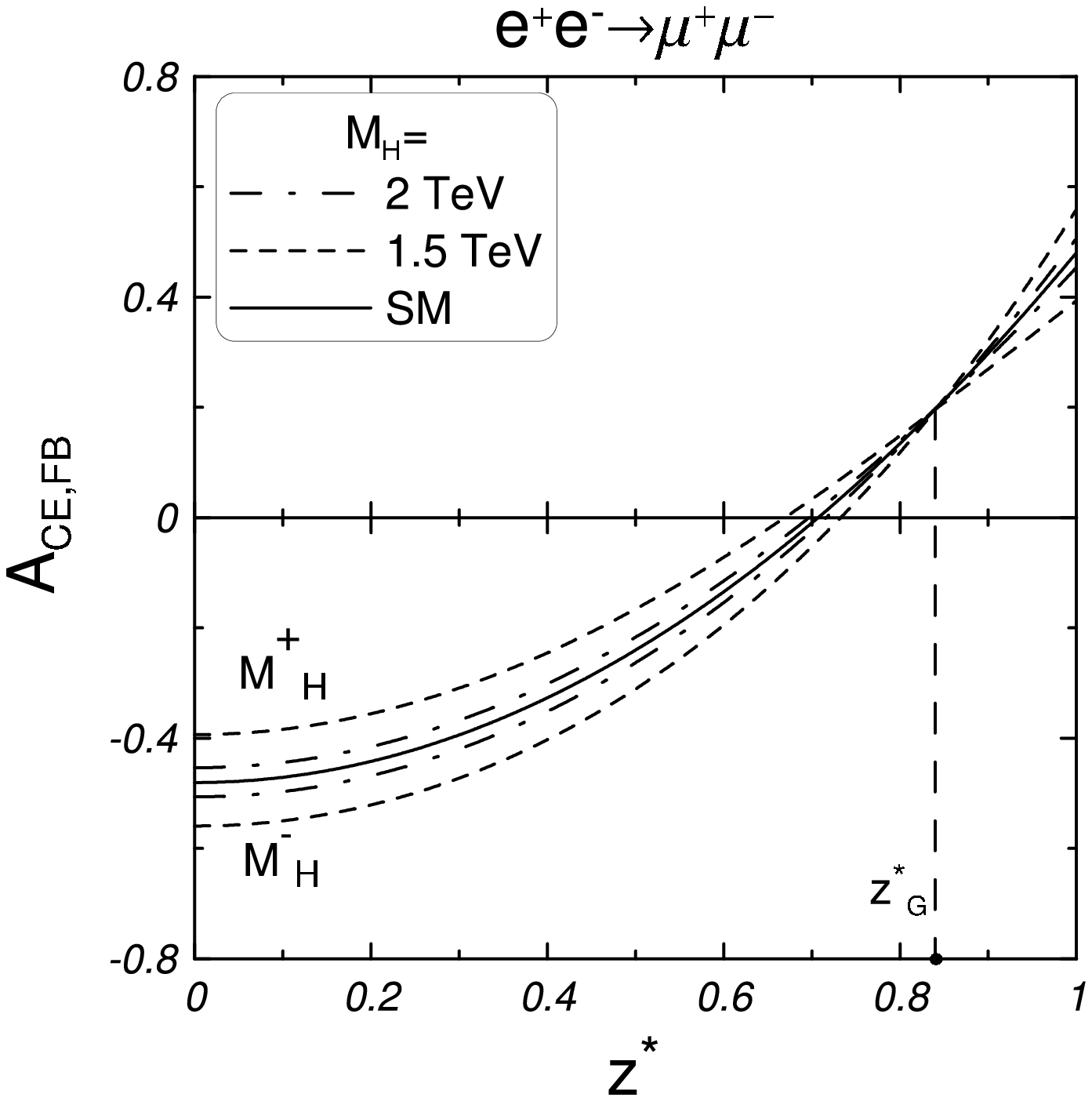,width=10.cm}
\vspace{-4.cm}
\caption{Center-edge-forward-backward asymmetry $A_{\rm CE,FB}$ as a function
of $z^*$ in the SM (solid line) and in the {\rm ADD} model with
$M_H^\pm=1$ TeV (dashed lines) and $M_H^\pm=1.5$ TeV (dash-dotted
lines). The $\pm $ superscripts correspond to constructive and distructive
interference of the SM amplitudes with the graviton exchange.
}
\end{center}
\end{figure}
\par
From the remarks above, we conclude that the measurement of
$A_{\rm CE,FB}(z^*_{\rm G})$ is sensitive only to new physics induced by
four-fermion contact interactions, free of contamination from graviton exchange
which does not contribute any deviation from the SM at that value of 
$z^*$. Therefore, contributions from CI interactions can unambiguously be 
identified. Conversely, as being not contaminated by contact interactions,
$A_{\rm CE,FB}(z^*_{\rm CI})$ is sensitive only to KK graviton
exchange, and can therefore be considered in combination with $A_{\rm CE}$
to improve the identification reach of the ADD scenario.

\begin{figure}[htb]
\begin{center}
\epsfig{file=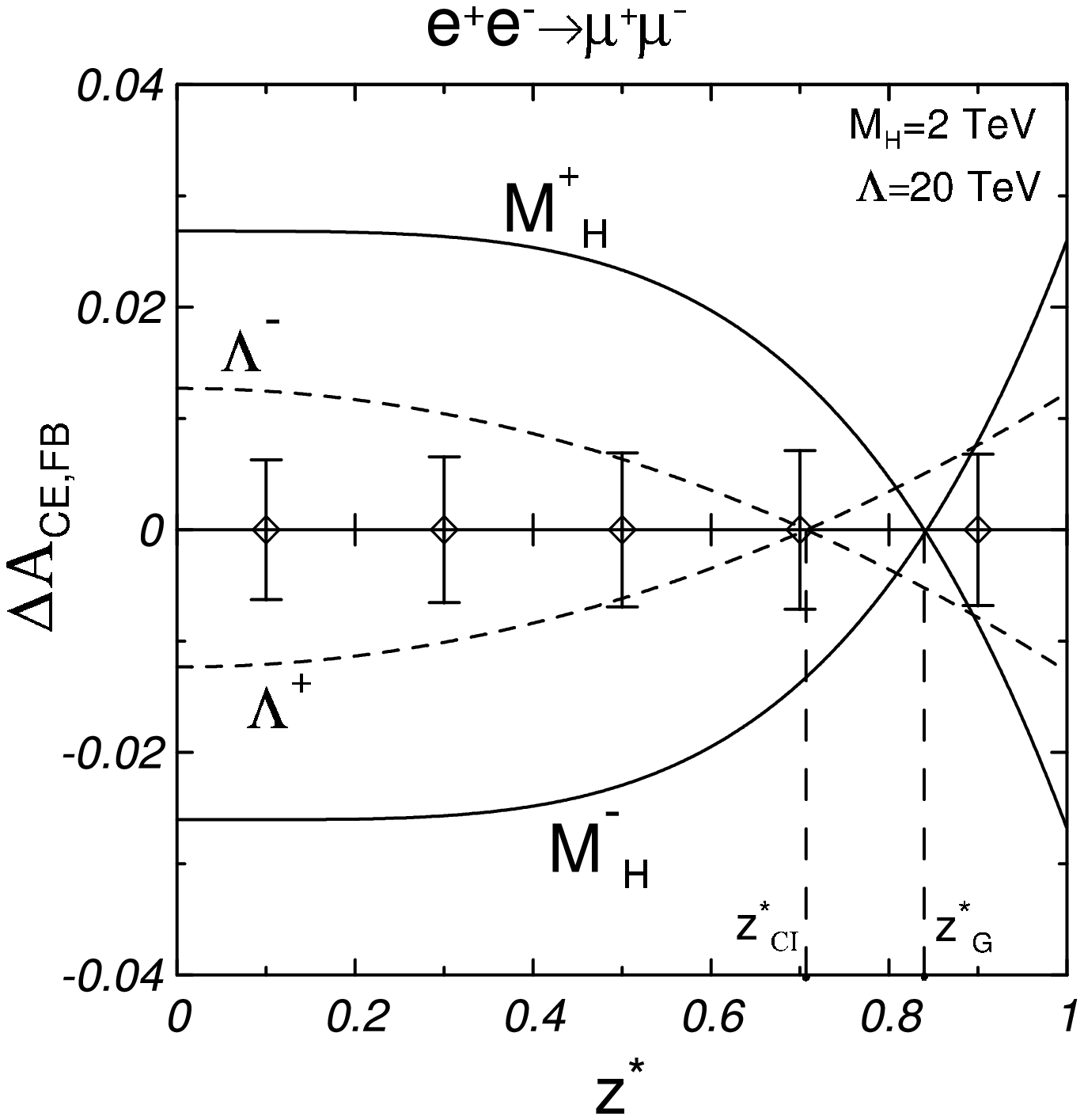,width=10.cm}
\vspace{-4.cm}
\caption{$\Delta A_{\rm CE,FB}(z^*)$ in the CI and the ADD scenarios
for the indicated values of $\Lambda$ and $M_H$. The $\pm$ superscripts 
refer to positive and negative interference, respectively. The vertical bars
represent the statistical uncertainty at a LC with $\sqrt s=0.5$ TeV and
${\cal L}_{\rm int}=50\, fb^{-1}$.
}
\end{center}
\end{figure}
\par
This suggests, in practice, the kind of analysis exemplified in Fig.~6,
where the deviations $\Delta A_{\rm CE,FB}(z^*)$ for the two scenarios
are compared to the statistical uncertainty expected at the Linear Collider.
Defining the sensitivity (or the signal statistical significance) to 
new physis as the ratio between the
deviation from the SM and the experimental uncertainty,
this figure shows that there is some range of $z^*$ around $z^*_{\rm CI}$
(and extending somewhat below this value) where
$\Delta A_{\rm CE,FB}^{\rm ADD}$ is appreciably larger than the
uncertainty, while $\Delta A_{\rm CE,FB}^{\rm CI}$ is still smaller than 
(or equal to) the uncertainty. Therefore, in this range, there is
maximal sensitivity to the KK graviton exchange. The converse is true
in a (limited) range of $z^*$ around $z^*_{\rm G}$, where the deviation
$\Delta A_{\rm CE,FB}^{\rm CI}$ can be dominant, and the sensitivity
to contact interactions will be much higher. In general, the widths of the
above mentioned $z^*$ intervals will be much larger than the expected
experimental uncertainty on $z$ which, therefore, will not affect
the numerical results presented in the following sections.

\section{Polarized beams}
In this case, with $P_1$ and $P_2$ the degrees of longitudinal polarization
of the electron and positron beams, respectively, we define
\cite{Flottmann:1995ga,Fujii:1995ys}
\begin{equation}
D=1-P_1 P_2\hskip 2pt , \qquad P_{\rm eff}=\frac{P_1-P_2}{1-P_1 P_2}.
\label{poleff}
\end{equation}
The polarized differential cross section can be expressed as follows:
\begin{equation}
\frac{\dd\sigma^{\rm pol}}{\dd z}=\frac{D}{4}
\left[(1-P_{\rm eff})
\left(\frac{\dd\sigma_{\rm LL}}{\dd z}
+\frac{\dd\sigma_{\rm LR}}{\dd z}\right)
+(1+P_{\rm eff})\left(\frac{\dd\sigma_{\rm RR}}{\dd z}
+\frac{\dd\sigma_{\rm RL}}{\dd z}\right)\right].
\label{crossdif-pol}
\end{equation}
The decomposition of the SM polarized differential cross section into
$z$-even and $z$-odd parts has identical structure as in Eq.~(\ref{dsigmasm}):
\begin{equation}
\frac{\dd\sigma^{\rm pol,SM}}{\dd z}=\frac{3}{8}\sigma^{\rm pol,SM}
\left(1+z^2\right)+\sigma_{\rm FB}^{\rm pol,SM}z,
\label{dsigmapolsm}
\end{equation}
and in this case, for the polarized forward-backward asymmetry,
Eq.(\ref{afbsm}) is replaced by the following one:
\begin{equation}
A_{\rm FB}^{\rm pol,SM}=\frac{3}{4}\hskip 2pt
\frac{\left(1-P_{\rm eff}\right)
\left[\left({\cal M}_{\rm LL}^{\rm SM}\right)^2  -
\left({\cal M}_{\rm LR}^{\rm SM}\right)^2\right] +
\left(1+P_{\rm eff}\right)
\left[\left({\cal M}_{\rm RR}^{\rm SM}\right)^2 -
\left({\cal M}_{\rm RL}^{\rm SM}\right)^2\right]}
{\left(1-P_{\rm eff}\right)
\left[\left({\cal M}_{\rm LL}^{\rm SM}\right)^2  +
\left({\cal M}_{\rm LR}^{\rm SM}\right)^2\right] +
\left(1+P_{\rm eff}\right)
\left[\left({\cal M}_{\rm RR}^{\rm SM}\right)^2 +
\left({\cal M}_{\rm RL}^{\rm SM}\right)^2\right]}.
\label{afbpolsm}
\end{equation}
\par
Using Eq.~(\ref{crossdif-pol}), one can define the polarized center--edge
asymmetry $A_{\rm CE}^{\rm pol}$ and the forward--backward--center--edge
asymmetry $A_{\rm CE,FB}^{\rm pol}$ in exactly the same way as in
Eqs.~(\ref{ace}), (\ref{sce}) and (\ref{acefb}), (\ref{scefb}), the only 
difference being that $\sigma$ must be replaced by $\sigma^{\rm pol}$
everywhere. Also, the same kind of decomposition into ``SM'',
``INT'' and ``NP'' contributions as in Eqs.~(\ref{acenp}) and (\ref{acefbnp})
can be written for the polarized asymmetries. One can easily see that the
$z^*$-dependence will remain the same as found in the previous sections,
while the structure of the $z^*$-independent factor expressed in terms of
the helicity amplitudes will be modified to account for the dependence on
the initial electron-positron spin configuration.

\subsection{The polarized center--edge asymmetry
{\boldmath $A_{\rm CE}^{\rm pol}$}}
Eqs.~(\ref{dsigmapolsm}) and (\ref{CI}) immediately show that
for the SM as well as for the CI the polarized center--edge asymmetry is
the same, and still given by Eq.~(\ref{acesm}) regardless of the c.m. energy 
and of the values of the beams longitudinal polarization. Therefore, the
same considerations made in Sect.~2.1 with regard to the unpolarized case
continue to hold, i.e., the center--edge asymmetry will be ``transparent''
to four-fermion contact interaction effects also for longitudinally 
polarized beams.
\par
Conversely, for the graviton exchage case, see Eq.~(\ref{ADD}), 
to first order in the coupling $f_G$ one has instead of Eq.~(\ref{deladd}):
\begin{align}
\Delta A_{\rm CE}^{\rm pol,ADD}(z^*)\cong&\frac{3}{4}f_G\hskip 2pt \frac{
\left(1-P_{\rm eff}\right)\left({\cal M}_{\rm LL}^{\rm SM} -
{\cal M}_{\rm LR}^{\rm SM}\right) + \left(1+P_{\rm eff}\right)
\left({\cal M}_{\rm RR}^{\rm SM} - {\cal M}_{\rm RL}^{\rm SM}\right)}
{\left(1-P_{\rm eff}\right)
\left[\left({\cal M}_{\rm LL}^{\rm SM}\right)^2  +
\left({\cal M}_{\rm LR}^{\rm SM}\right)^2\right] +
\left(1+P_{\rm eff}\right)
\left[\left({\cal M}_{\rm RR}^{\rm SM}\right)^2 +
\left({\cal M}_{\rm RL}^{\rm SM}\right)^2\right]}
\nonumber  \\
&\times 4z^*\left(1-{z^*}^2\right).
\label{acepoladd}
\end{align}

\subsection{The polarized asymmetry {\boldmath $A_{\rm CE,FB}^{\rm pol}$}}
For this observable, the $z$-integration of the differential cross section
(\ref{crossdif-pol}) is the same as in Eqs.~(\ref{acefb}) and (\ref{scefb}).
Of course, also in this case a separation between SM and NP effects analogous
to Eqs.~(\ref{acenp}) and (\ref{acefbnp}) holds. Similar to Sect.~3.1,
the polarized center-edge-forward-backward asymmetries will have the same
$z^*$-dependence as in the unpolarized case, times a $z^*$-independent
factor accounting for the initial beams polarization configuration.
\par
Thus, using the structure of Eqs.~(\ref{crossdif-pol}) and (\ref{CI}), one 
finds for the SM and for the contact interactions cases, respectively:
\begin{equation}
A_{\rm CE,FB}^{\rm pol, SM}(z^*)=A_{\rm FB}^{\rm pol,SM}\left(2{z^*}^2-1\right)
\qquad {\rm and}
\qquad
A_{\rm CE,FB}^{\rm pol,CI}(z^*)=
A_{\rm FB}^{\rm pol,CI}\left(2{z^*}^2-1\right),
\label{acefbpolci}
\end{equation}
so that the deviation from the SM is given by
\begin{equation}
\Delta A_{\rm CE,FB}^{\rm pol,CI}(z^*)=
\Delta A_{\rm FB}^{\rm pol,CI}\left(2{z^*}^2-1\right).
\label{deviat-acefbpol-ci}
\end{equation}
\par
For graviton exchange, to first order in $f_G$ one finds the relations:
\begin{equation}
A_{\rm CE,FB}^{\rm pol,ADD}(z^*)=
A_{\rm FB}^{\rm pol,SM}\left(2{z^*}^2-1\right)+
\Delta A_{\rm FB}^{\rm pol,ADD}\left(2{z^*}^4-1\right),
\label{acefbpoladd}
\end{equation}
or, for the deviation from the SM:
\begin{equation}
\Delta A_{\rm CE,FB}^{\rm pol,ADD}(z^*)\cong
\Delta_{\rm FB}^{\rm pol,ADD}\left(2{z^*}^4-1\right),
\label{deviat-acefbpol-add}
\end{equation}
with
\begin{equation}
\Delta A_{\rm FB}^{\rm pol,ADD}\cong
-\frac{3}{4}f_G\hskip 2pt
\frac{\left(1-P_{\rm eff}\right)\left({\cal M}_{\rm LL}^{\rm SM} +
{\cal M}_{\rm LR}^{\rm SM}\right) + \left(1+P_{\rm eff}\right)
\left({\cal M}_{\rm RR}^{\rm SM}+{\cal M}_{\rm RL}^{\rm SM}\right)}
{\left(1-P_{\rm eff}\right)
\left[\left({\cal M}_{\rm LL}^{\rm SM}\right)^2  +
\left({\cal M}_{\rm LR}^{\rm SM}\right)^2\right] +
\left(1+P_{\rm eff}\right)
\left[\left({\cal M}_{\rm RR}^{\rm SM}\right)^2 +
\left({\cal M}_{\rm RL}^{\rm SM}\right)^2\right]}.
\label{deviat-afbpol-add}
\end{equation}

\section{Identification reaches on {\boldmath $\Lambda$}
and {\boldmath $M_H$}}
Essentially, to asses the {\it identification} reach on the mass scales 
$M_H$ and $\Lambda$ at the Linear Collider, we can compare the 
deviations from the SM predictions of the asymmetries defined in the 
previous sections with the expected experimental uncertainties on these 
observables. This kind of analysis is based on a $\chi^2$ function, 
definded as
\begin{equation}
\chi^2=\frac{\left(\Delta O^f\right)^2}{\left(\delta O^f\right)^2},
\label{chisquare}
\end{equation}
where the superscript ``$f$'' refers to the final state in process
(\ref{proc}); $O^f$ indicate our observables for the considered
final state, i.e., $O=A_{\rm CE}$ and $A_{\rm CE,FB}$; $\Delta O$
indicate the deviations from the SM, whose explicit expressions are reported
in Sects.~2.1 and 2.2 (and Sects.~3.1 and 3.2 in the case of
polarized $e^+e^-$ beams); finally, $\delta O$ are the
corresponding expected experimental uncertainties. In case, total  
(or partial) summation of the right-hand-side of Eq.~(\ref{chisquare}) over   
the final states $f=\mu^+\mu^-$, $\tau^+\tau^-$, $c\bar c$ and $b\bar b$ 
considered here may be performed to derive combined constraints on 
$M_H$ and $\Lambda$. Basically, numerical constraints on the new physics 
parameters follow from the condition
\begin{equation}
\chi^2\leq\chi^2_{\rm crit},
\label{chicrit}
\end{equation}
where the actual numerical value of $\chi^2_{\rm crit}$ depends on the
desired confidence level (C.L.).
\par
Experimental uncertainties are determined by the combination of statistical 
uncertainties, depending on the Linear Collider integrated luminosity and 
of systematic uncertainties reflecting experimental details. Since the
above mentioned asymmetries are basically ratios of integrated cross
sections, one expects systematic errors to cancel to a very large
extent and, indeed, the uncertainty turns out to be numerically dominated 
by the statistical one and by the uncertainty on initial beams polarization. On
the other hand, as mentioned in Sect.~2, the deviations from the SM increase
with the c.m. energy $\sqrt s$. Therefore, searches on $M_H$ and 
$\Lambda$ are favoured by the high energies and the high luminosities 
in $e^+e^-$ collisions envisaged at the planned Linear 
Collider \cite{Aguilar-Saavedra:2001rg}.
\par
Specifically, in our numerical analysis we consider a LC with 
$\sqrt s=0.5$ TeV and $1$ TeV, to assess the dependence of the results 
on the c.m. energy, and time-integrated luminosity ${\cal L}_{\rm int}$
ranging from 50 up to 1000 ${\rm fb}^{-1}$. 
As far as uncertainties are concerned, we assume 
$\Delta {\cal L}_{\rm int}/{\cal L}_{\rm int}
= \Delta P_1/P_1=\Delta P_2/P_2 =0.5\%$,
and polarizations $\vert P_1\vert=80\%$ and $\vert P_2\vert=60\%$ for
electron and positron beams, respectively. Also, a realistic value
that we assume in our analysis is the angular resolution 
$\Delta\theta=0.5$ mrad. In all cases, a small angle cut of $10^\circ$ 
around the beam pipe has been assumed (the results are found not 
particularly sensitive to the value of this cut).
\par
Regarding the theoretical inputs, for the SM amplitudes 
we use the effective Born approximation \cite{Consoli:1989pc} taking into 
account electroweak corrections to the propagators and vertices, 
with $m_{\rm top}=175~\text{GeV}$ and $m_{\rm higgs}=300~\text{GeV}$. 
Also, ${\cal O}(\alpha)$ corrections to process (\ref{proc}) 
are taken into account, along the lines followed in 
Ref.~\cite{Osland:2003fn}. Basically, the numerically most important 
QED corrections, from initial-state radiation, are calculated in the flux 
function approach (see, {\it e.g.}, Ref.~\cite{physicsatlep2}). To minimize 
the effect of radiative flux return to the $Z$ whereby the emitted 
photons peak in the ``hard'' region 
$E_\gamma/E_{\rm beam}\approx 1-M^2_{Z}/s$, and thus to increase the chances 
for new physics signals, we apply a cut on the radiated photon energy 
$\Delta=E_\gamma/E_{\rm beam}< 0.9$. As far as the final-state and 
initial-final state corrections are concerned, they are evaluated by 
using ZFITTER \cite{Bardin:2001yd} and found to be unimportant, for the 
chosen values of the kinematical cuts, in the derivation of the final 
numerical results on $M_H$ and $\Lambda$. Moreover, the positions of the 
zeros of the basic asymmetries, $z^*_0$, $z^*_{\rm CI}$ and $z^*_{\rm G}$ 
(see Eqs.~(\ref{z*0}), (\ref{z*CI}) and (\ref{z*G})), can be shifted by the 
above mentioned QED corrections by a small amount, such that the results of 
the analysis are practically unaffected.\footnote{This is true also of the 
box-diagram contributions, which introduce different angular dependences.}

\subsection{Limits on graviton exchange}
In Fig.~7, we show the 5$\sigma$ {\it identification} reach on the
graviton exchange mass scale $M_H$ as a function of luminosity and
of c.m. energy, obtained from the conventional $\chi^2$ analysis
combining $A_{\rm CE}$ and $A_{\rm CE,FB}$ at $z^*=z^*_{\rm CI}$.
Also, three possible initial beams longitudinal polarization
configurations have been considered in this figure. In particular,
for $A_{\rm CE}$ ($A_{\rm CE,FB}$) we take $P_1=P_2=0$, $P_1=0.8$,
$P_2=0$ and $P_1=0.8$ $P_2=-0.6$ ($P_1=P_2=0$, $P_1=-0.8$, $P_2=0$
and $P_1=-0.8$, $P_2=0.6$) for the cases unpolarized beams, polarized
electrons, and both beams polarized, as such values of longitudinal 
polarizations are numerically found to provide the maximal sensitivity of 
the asymmetries to $M_H$. In all three cases the difference between the 
results for positive and negative interference 
($\lambda=\pm 1$ in Eq.~(\ref{ADD})) are small and cannot be made visible 
on the scale of the figure.
\begin{figure}[htb]
\begin{center}
\epsfig{file=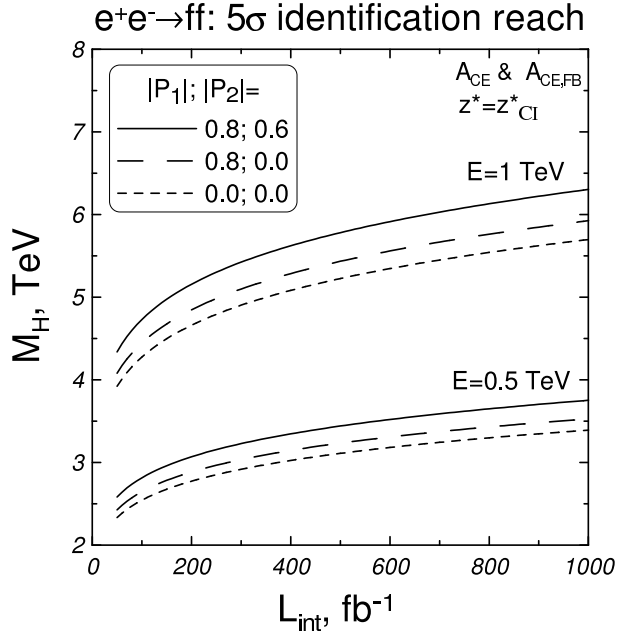,width=10.cm} \vspace{-4.cm}
\caption{5$\sigma$ identification reach on the mass scale $M_H$
{\it vs.}\ integrated luminosity obtained from the combined
analysis of two polarized asymmetries, $A_{\rm CE}$ and $A_{\rm
CE,FB}$, for the process $e^+e^-\to f{\bar f}$, with $f$ summed
over $\mu,\tau,b,c$, at $z^*=z^*_{\rm CI}$ and at the c.m. energy
of 0.5~TeV and 1~TeV. Short-dashed: unpolarized; long-dashed:
polarized electrons, $|P_1|=0.8$; solid: both beams polarized,
$|P_1|=0.8$, $|{P_2}|=0.6$. }
\end{center}
\end{figure}
\par
As one can see, the dependence of the reach on $M_H$ on the
luminosity is rather smooth (dimensionally, including the
statistical error only, we would expect the bound on $M_H$ to
scale like $\sim\left({\cal L}_{\rm int}s^3\right)^{1/8}$). Also,
electron and positron longitudinal polarizations can contribute a
significant improvement in the sensitivity to graviton exchange,
but, at fixed c.m. energy and luminosity, the impact on $M_H$ is not 
so dramatic due to the high power of the suppression factor 
$\sim\sqrt s/M_H$ in Eq.~(\ref{ADD}), reflecting the dimension-8 relevant 
operator of Eq.~(\ref{dim-8}). This has to be compared with the case of 
contact interactions, see Eq.~(\ref{CI}), where the dependence on the 
suppression factor $\sqrt s/\Lambda$ is only quadratic. Also, retaining 
the statistical uncertainty only, the bound on $\Lambda$ would scale 
like $\sim \left(s{\cal L}_{\rm int}\right)^{1/4}$.
\par
It should be interesting to make a comparison of the {\it identification 
reach} derived at the LC using the kind of analysis described above, with 
the results on $M_H$ potentially obtainable from the study of the inclusive 
di-lepton production process 
\begin{equation}
p+p\to l^+ l^- + X,
\label{drellyan}
\end{equation}
using the observable $A_{\rm CE}$ at the proton-proton collider LHC 
\cite{Dvergsnes:2004tw}. This comparison is performed in Fig.~8, which shows 
the identification reaches at the 95\% C.L. obtainable at the LC for various 
c.m. energies and luminosities, and at the LHC with 
${\cal L}_{\rm int}=100\hskip 2pt {\rm fb}^{-1}$. This figure suggests that 
the LC(0.5 TeV) can be competitive to the LHC for 
${\cal L}_{\rm int}\ge 500\hskip 2pt {\rm fb}^{-1}$, whereas the 
LC(1 TeV) is definitely superior for any value of the luminosity.
\begin{figure}[htb]
\begin{center}
\epsfig{file=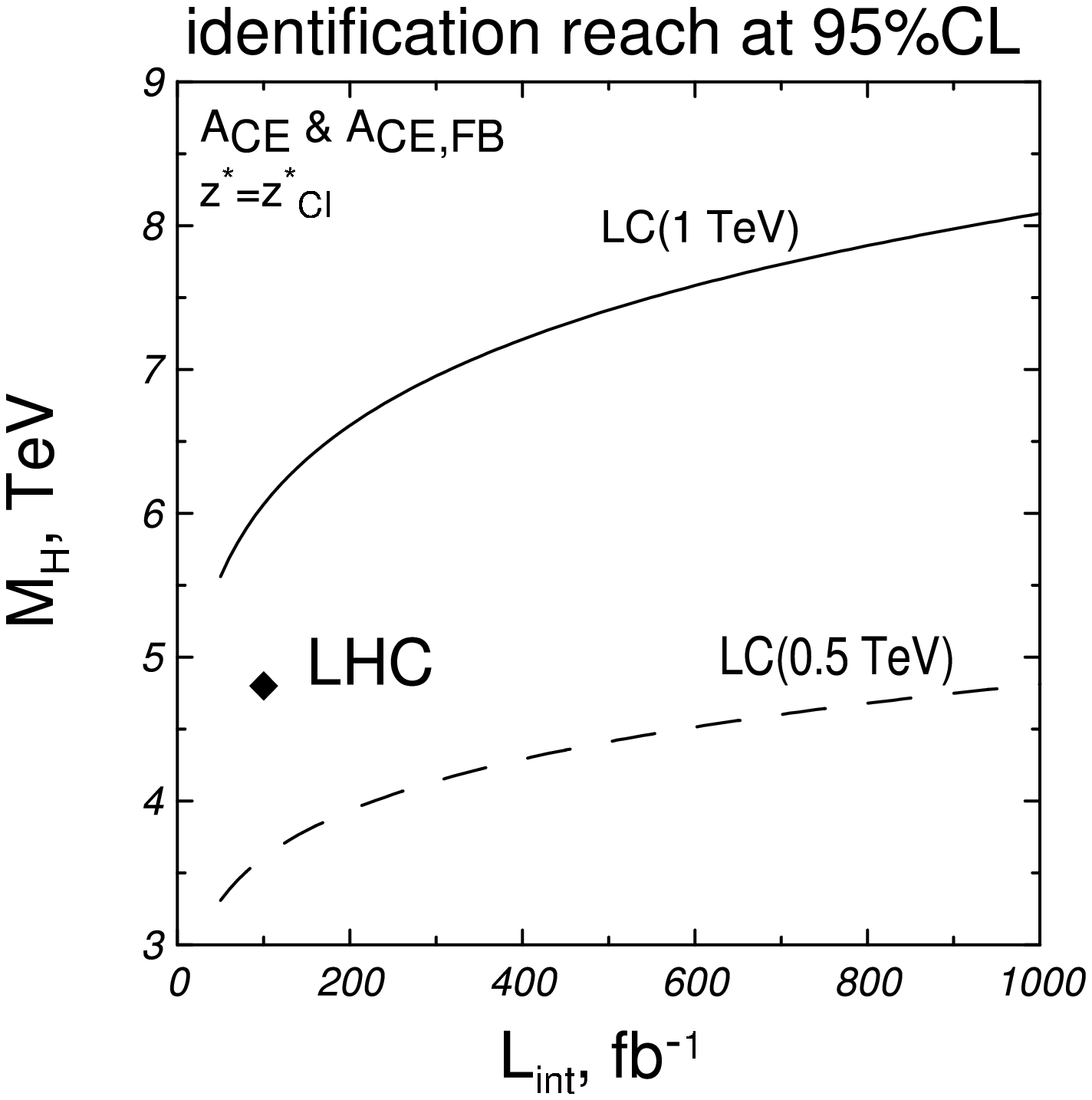,width=10.cm} \vspace{-4.cm} \caption{95\% C.L. 
identification  reach on the mass scale $M_H$ vs. integrated
luminosity from the dilepton production process at LHC using the 
center-edge asymmetry and from the processes $e^+e^-\to\bar{f}f$
at LC for the c.m. energy $\sqrt{s}=0.5$ TeV and 1 TeV combining 
$A_{\rm CE}$ and $A_{\rm CE,FB}$ as described in Fig.~7.}
\end{center}
\end{figure}
\par
As repeatedly stressed in Sects. 3.1 and 3.2, the basic observables 
$A_{\rm CE}$ and $A_{\rm CE,FB}(z^*_{\rm CI})$  have the feature that  
deviations from the SM predictions, if experimentally observed at the LC 
within the expected accuracy, can be unambiguously associated to graviton 
exchange.

\subsection{Limits on contact interactions}
We now consider the reach on the four-fermion contact-interaction  
scales $\Lambda$ defined in Eq.~(\ref{CI}), obtainable at the Linear 
Collider. In Fig.~9, we report the 5$\sigma$ 
reach as a function of the time-integrated luminosity and for 
two options for the c.m. energy, obtained by applying the $\chi^2$ analysis 
described above to $A_{\rm CE,FB}^{\rm pol}$ at $z^*=z^*_G$, and 
combining the final $\mu^+\mu^-$ and $\tau^+\tau^-$ channels. Recall that, 
as pointed out in Sect.~3.1, the asymmetry $A_{\rm CE}$ is not sensitive to 
contact interactions, because the deviation from the SM vanishes.  
In Fig.~9, the longitudinal polarizations of the initial $e^-$ and $e^+$ 
beams leading to maximal sensitivity are specified in the caption. 
The four curves in each of the two panels are obtained by assuming non-zero 
values for only one of the $\eta_{\alpha\beta}$ configurations 
of Eq.~(\ref{lagra}) at a time, and all the others equal to zero 
(one-parameter fit). As anticipated, the increase of 
$\Lambda$s with ${\cal L}_{\rm int}$ is much steeper compared to 
the case of $M_H$, reflecting the dimension-6 of the relevant operators.
\begin{figure}[htb]
\refstepcounter{figure} \label{Fig:fig9} \addtocounter{figure}{-1}
\begin{center}
\setlength{\unitlength}{1cm}
\begin{picture}(10.0,10.0)
\put(-4,-0.5) {\mbox{\epsfysize=10.cm\epsffile{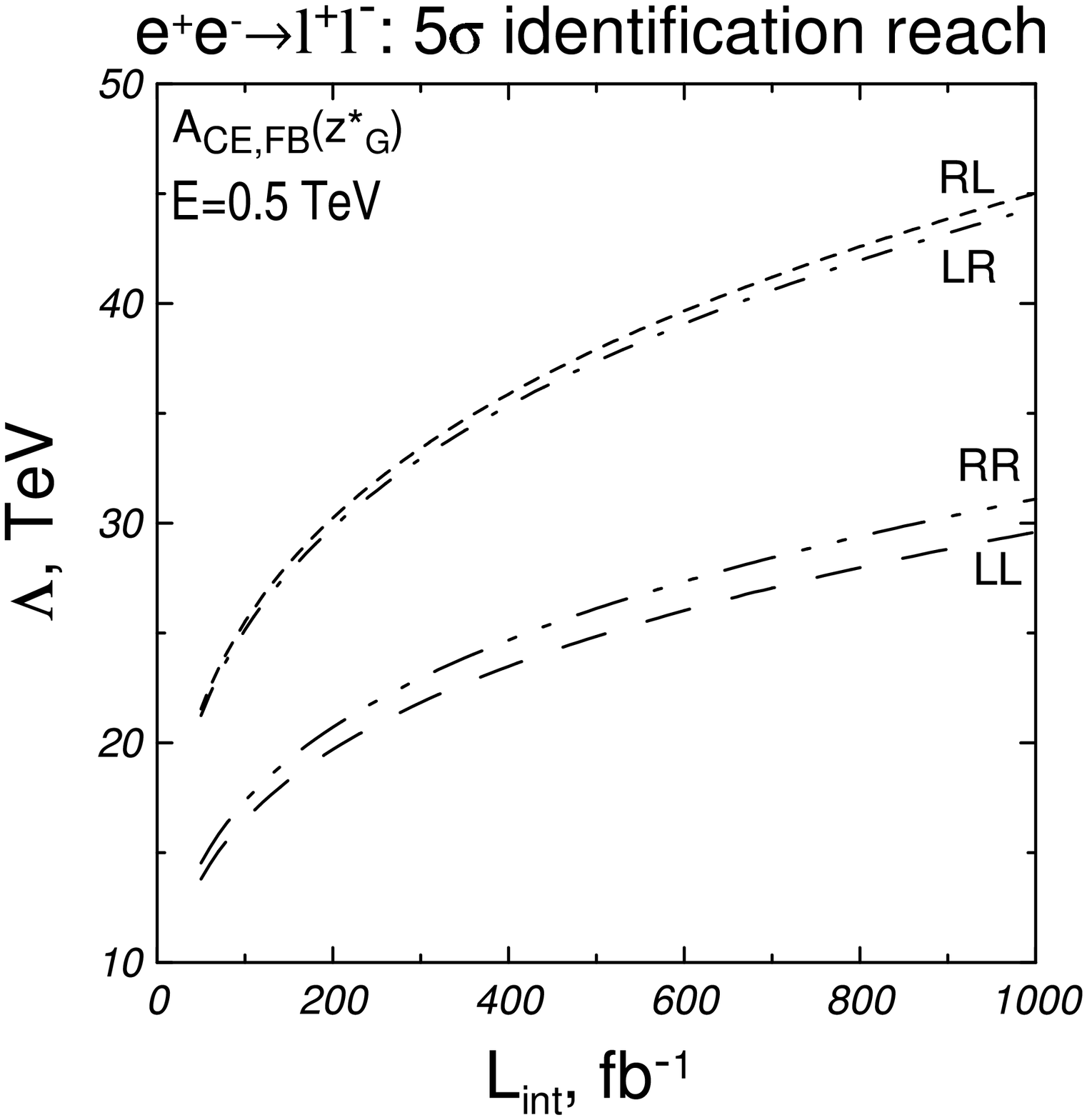}}
 \mbox{\epsfysize=10.cm\epsffile{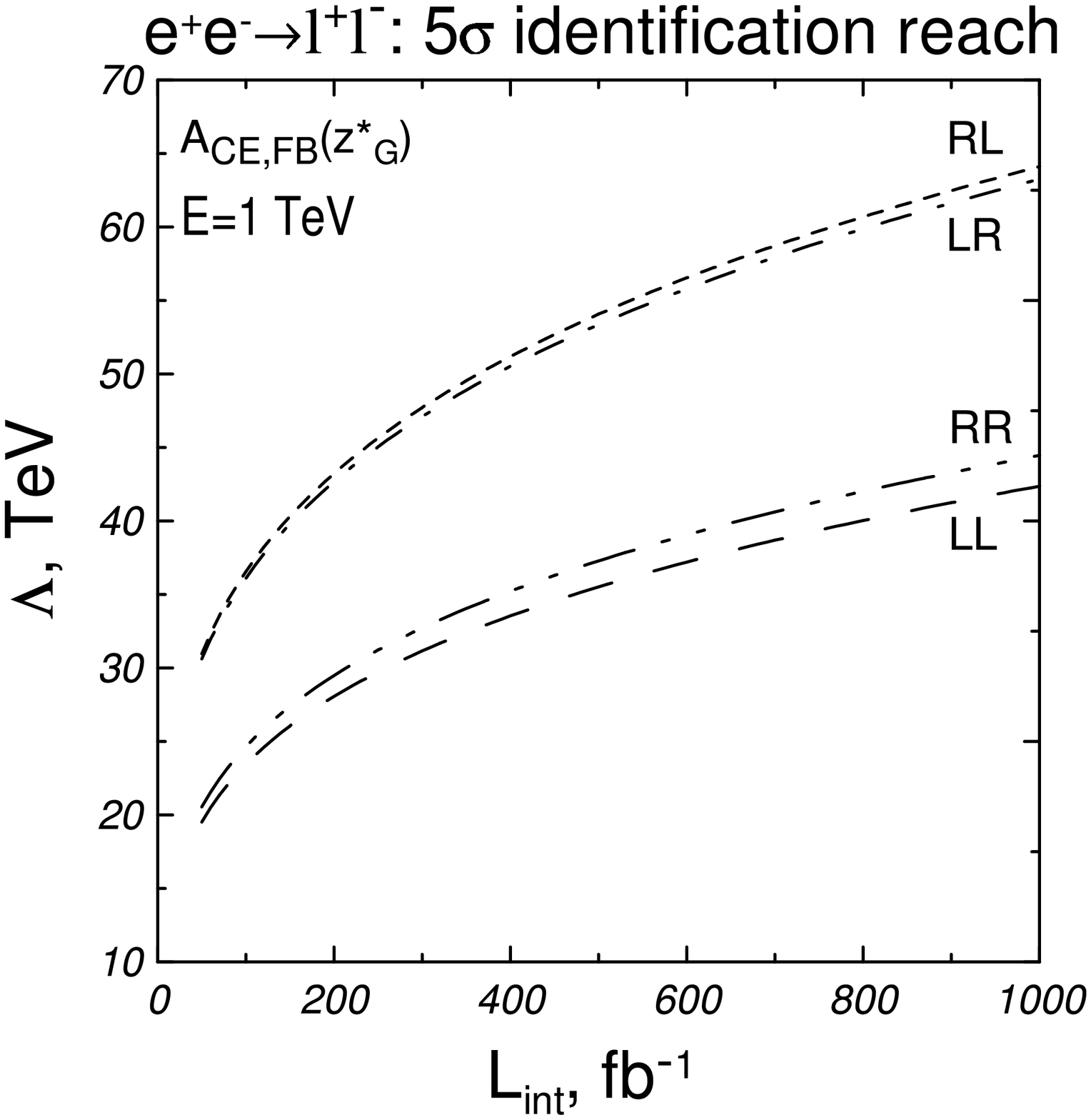}}}
\end{picture}
\vspace*{-4mm} 
\caption{5$\sigma$ reach on the mass
scales $\Lambda_{\alpha\beta}$ {\it vs.}\ integrated luminosity
obtained from the one-parameter fit of polarized
center-edge-forward-backward asymmetry $A_{\rm CE,FB}^{\rm pol}(z^*_{\rm G})$
for the process $e^+e^-\to l^+l^-$, with $l$ summed over
$\mu,\tau$, at the c.m. energy $\sqrt{s}=0.5$
TeV (left panel) and 1 TeV (right panel) and $P_1=0.8$,
$P_2=-0.6$. Labels attached to the curves indicate the helicity 
configurations ($\alpha\beta={\rm LL,RR,LR,RL}$).
}
\end{center}
\end{figure}
\par 
As discussed in Sect.~3.2, the values of $\Lambda_{\alpha\beta}$ reported in 
Fig.~9 should not be contaminated from effects of graviton exchange, that 
vanish in $A_{\rm CE,FB}$ at $z^*=z^*_{\rm G}$. On the other hand, 
as previously remarked, one cannot avoid competitive virtual effects from 
other kinds of effective 4-fermion interactions, in our case, the chosen 
example of a very heavy sneutrino exchange in the $t$-channel, see 
Eq.~(\ref{nutilde}). For an illustration of this effect, in Fig.~10 we 
report the bounds in the ($\lambda-{m}$) plane that would obtain by 
considering deviations from the SM of $A_{\rm CE}$ at $z^*=z^*_0$ and of  
$A_{\rm CE,FB}$ at $z^*=z^*_{\rm CI}$ generated from Eq.~(\ref{nutilde}). 
Here, the c.m. energy is $\sqrt s=0.5$ TeV and the integrated luminosity 
is ${\cal L}_{\rm int}=50$ and 500 ${\rm fb}^{-1}$.\footnote{We consider 
only leptonic final states, separately because universality is not 
automatically assured in this framework.}
\begin{figure}[htb]
\refstepcounter{figure} \label{Fig:fig10} \addtocounter{figure}{-1}
\begin{center}
\setlength{\unitlength}{1cm}
\begin{picture}(10.0,10.0)
\put(-4,-0.5) {\mbox{\epsfysize=10.5cm\epsffile{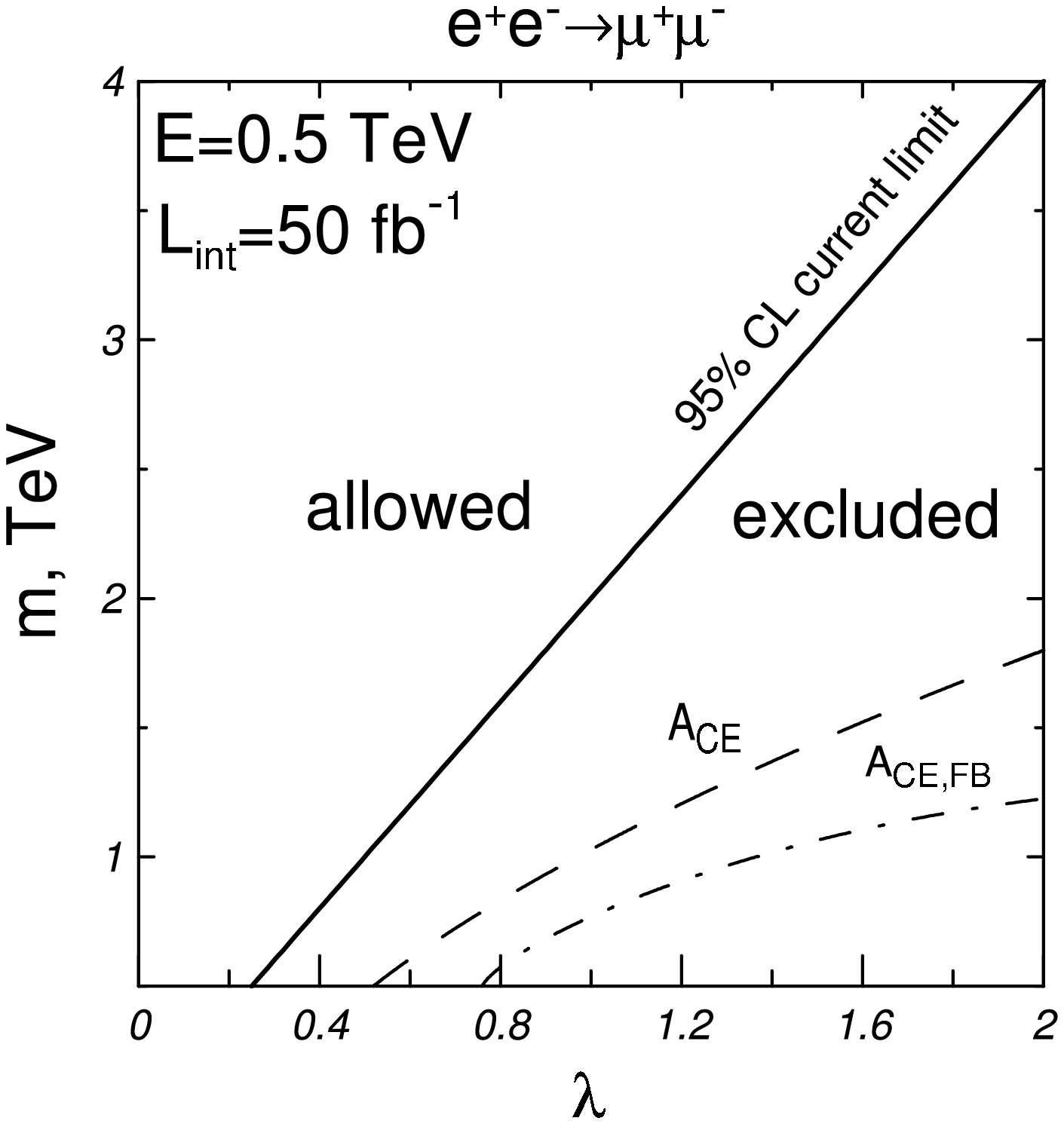}}
\mbox{\epsfysize=10.5cm\epsffile{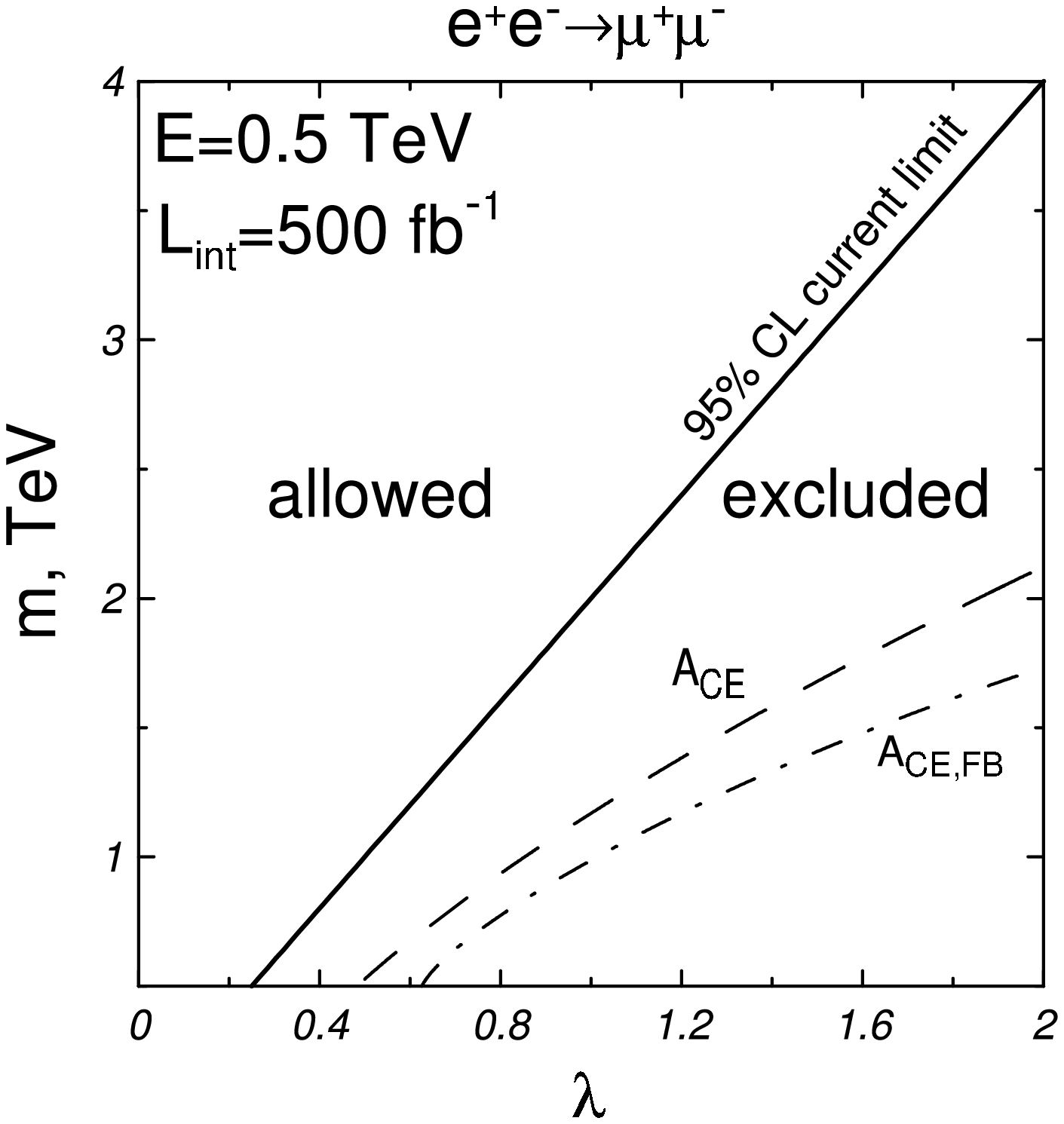}}}
\end{picture}\vspace*{-2.4cm} 
\caption{
Indirect search reach at 95\% C.L. for $t$-channel
exchange sneutrino $\tilde\nu$ as a function of its mass from the
process $e^+e^-\to\mu^+\mu^-$ at $\sqrt{s}=0.5$ TeV with
integrated luminosity $\mathcal{L}_{int}=50\ fb^{-1}$ (left panel) and  
$500\ fb^{-1}$ (right panel) using $A_{\rm CE,FB}$ at $z^*=z^*_{\rm CI}$, 
$P_1=-0.8$, $P_2=0.6$ and $A_{\rm
CE}$ at $z^*=z^*_0$, $P_1=0.8$, $P_2=-0.6$.
}
\end{center}
\end{figure}

In Fig.~10, the full straight line represents the current limit derived from 
low-energy physics in Ref.~\cite{Kalinowski:1997bc}. As one can see, the 
curves labelled as $A_{\rm CE}$ and $A_{\rm CE,FB}$ fall far-below the 
current limit, i.e., the  
observables $A_{\rm CE}(z^*_0)$ and $A_{\rm CE,FB}(z^*_{\rm CI})$ are not 
sensitive to $\tilde\nu$-exchange. Indeed, the $z$-independent 
contact-interaction limit of Eq.~(\ref{nutilde}) cannot contribute to 
$A_{\rm CE}$ and $A_{\rm CE,FB}$ at the above mentioned respective values of 
$z^*$, and only the remaining, strongly suppressed, $t$-dependent part would 
remain to give a non-zero contribution. One therefore can conclude that 
sneutrino exchange should not contaminate the limits on the graviton scale 
parameter $M_H$ derived in Sect.~5.1. 
\par
On the other hand, in Fig.~11 we report 
the bounds in the ($\lambda-{m}$) plane that one would obtain by 
assuming deviations of $A_{\rm CE,FB}(z^*_{\rm G})$ to be generated by 
Eq.~(\ref{nutilde}) instead of (\ref{CI}). The figure indicates that 
at $z^*_{\rm G}$ one cannot unambiguously distinguish the two kinds of 
new physics, vector-vector contact interactions from sneutrino exchange, so 
that in the case of four-fermion contact interaction one more 
appropriately should speak of  {\it discovery reach} rather 
than {\it identification reach}. On the other hand, the figure shows an 
interesting possibility to substantially extend the constraints on the 
sneutrino parameters.  
\begin{figure} [htb]
\begin{center}
\epsfig{file=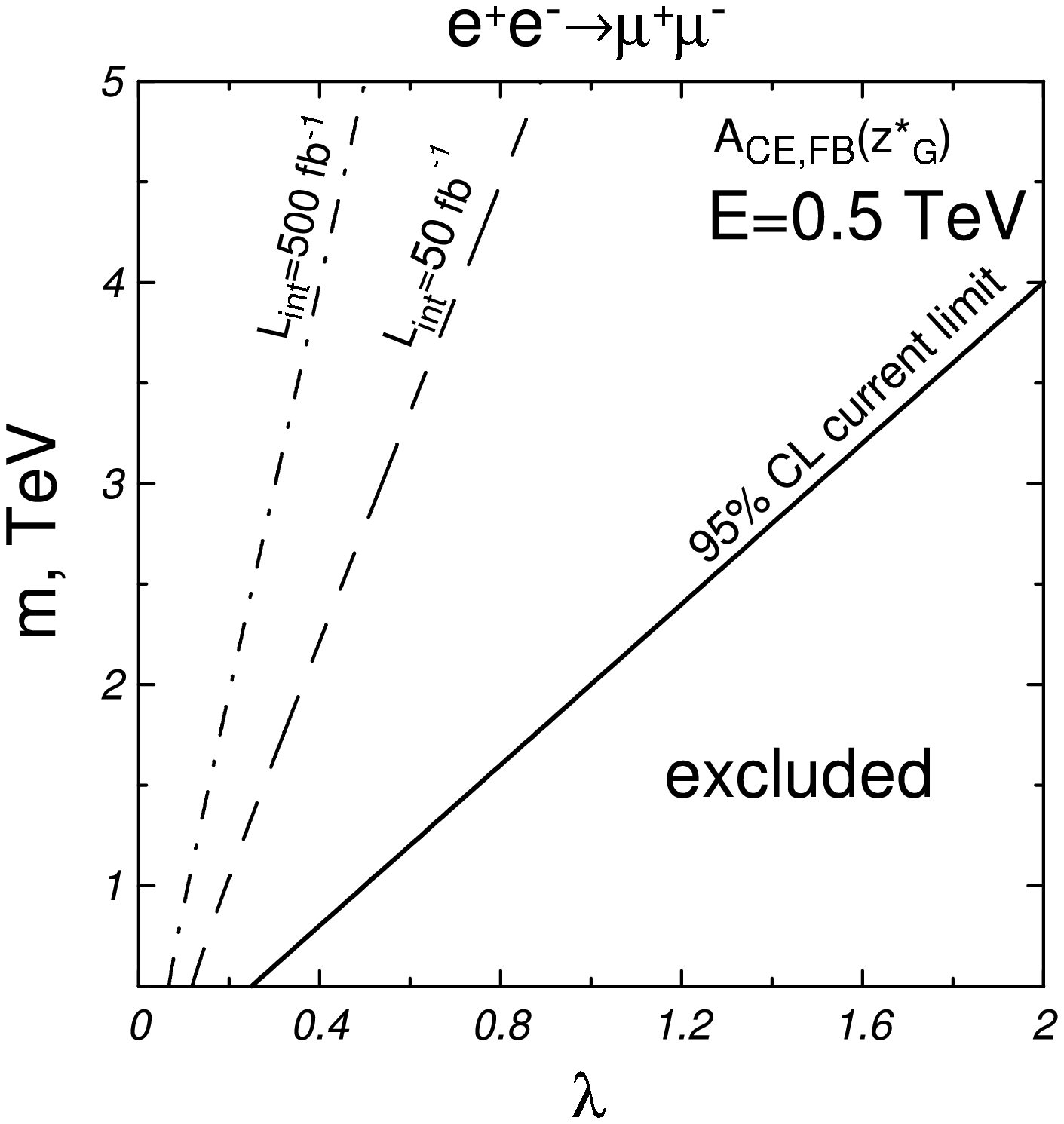,width=10.cm} \vspace{-4.cm}
\caption{Indirect search reach at 95\% C.L. for $t$-channel
exchange sneutrino $\tilde\nu$ as a function of its mass from the
process $e^+e^-\to\mu^+\mu^-$ at $\sqrt{s}=0.5$ TeV with
integrated luminosity $\mathcal{L}_{int}=50\ fb^{-1}$ (dashed line) and
$500\ fb^{-1}$ (dot-dashed line) using $A_{\rm CE,FB}$ 
at $z^*=z^*_{\rm G}$ with $P_1=-0.8$, $P_2=0.6$.
}
\end{center}
\end{figure}
\section{Concluding remarks}
In the previous sections, we have discussed the possible uses of center-edge 
asymmetries to pin down spin-2 graviton exchange signatures in the framework 
of large extra dimensions provided by the ADD model, and our findings can 
be summarized as follows.
\par 
The interference of SM and KK graviton exchanges in process (\ref{proc}) 
produces both $\cos\theta$-even and $\cos\theta$-odd contributions to the 
angular distribution of outgoing fermions. The appearance of such even and 
odd new terms does not occur in the case of other new physics, such 
as the four-fermion contact interactions considered here, and also in some 
different versions of the extra dimensions framework, for instance the 
one relevant to gauge boson KK excitations. 
These interference effects can be directly probed by the center-edge 
asymmetries, the even ones by $A_{\rm CE}$ and the odd ones by 
$A_{\rm CE,FB}$, providing uniquely distinct signatures.
\par
Specifically, the ``even'' center-edge asymmetry $A_{\rm CE}$ is sensitive 
{\it only} to the KK graviton exchange within almost the whole range of 
the angular kinematical parameter $z^*$ used in its definition. 
Conversely, the ``odd'' center-edge-forward-backward asymmetry 
$A_{\rm CE,FB}$ is able to project out either conventional four-fermion 
contact interaction effects or KK the graviton exchange ones by 
choosing appropriate values of $z^*$. In particular, $A_{\rm CE,FB}$ is 
not affected by spin-2/graviton exchange or by four-fermion 
contact interactions at $z^*=z^*_{\rm G}\simeq 0.841$ and at 
$z^*=z^*_{\rm CI}\simeq 0.707$, respectively. 
Accordingly, $A_{\rm CE,FB}$ can be used to study the identification reach 
of both the graviton exchange and the conventional contact interactions.
\par
As regards the numerical limits on the cut-off $M_H$, using the 
combination of $A_{\rm CE}$ and with the relevant $A_{\rm CE,FB}$, 
it is possible to select KK graviton exchange without contamination from 
the other new physics at the $5\sigma$ level up to values of 
$M_H\sim (6.3-7.5)\sqrt{s}$, that represent a substantial improvement over 
the current situation, also considering that this is an {\it identification} 
reach.  
\par
Using this same kind of analysis for the leptonic processes 
$e^+e^-\to l^+l^-$, we obtain the $5\sigma$ discovery reach 
for the contact interaction mass scales $\Lambda$ which range up to 45 TeV 
and 65 TeV for c.m. en\-er\-gies $\sqrt{s}=0.5$ TeV and 1 TeV, 
re\-spec\-tively.
\bigskip
\bigskip
\bigskip
\par\noindent 
\leftline{\bf Acknowledgements}
\par\noindent
AAP acknowledges the support of INFN in the early stage of this work. 
NP has been partially supported by funds of MIUR (Italian Ministry of 
University and Research) and of the University of Trieste. 
\medskip

\end{document}